\documentclass[12pt]{article}
\usepackage{latexsym}

\input epsf
\def\t{\widetilde}
\def\beq{\begin{eqnarray}}
\def\eeq{\end{eqnarray}}

\def\mpl{M_{\rm Pl}}

\def\lsim{\mathrel{\rlap{\lower3pt\hbox{\hskip0pt$\sim$}}
     \raise1pt\hbox{$<$}}}         
\def\gsim{\mathrel{\rlap{\lower4pt\hbox{\hskip1pt$\sim$}}
     \raise1pt\hbox{$>$}}}         

\begin{document}


\begin{flushright}
{NYU-TH-06/03/9}
\end{flushright}
\vskip 0.9cm

\centerline{\Large \bf 
(De)coupling Limit of DGP}
\vspace{0.5cm}

\vskip 0.7cm
\centerline{\large Gregory Gabadadze and  Alberto Iglesias}
\vskip 0.3cm
\centerline{\em Center for Cosmology and Particle Physics}
\centerline{\em Department of Physics, New York University, New York, 
NY, 10003, USA}

\vskip 1.9cm 

\begin{abstract}

We investigate the decoupling limit in the DGP model of gravity
by studying its nonlinear equations of motion.  We show that,
unlike 4D massive gravity, the limiting theory 
does not reduce to a sigma model of a single scalar field: 
Non-linear mixing terms of the scalar with a tensor also 
survive. Because of these terms physics of DGP is  different from that
of the scalar  sigma model. We show that 
the static spherically-symmetric solution of the scalar model 
found in  hep-th/0404159, is not a solution of the 
full set of nonlinear equations. As a consequence of this, 
the interesting result on hidden superluminality uncovered 
recently in the scalar model in  hep-th/0602178, 
is not applicable to the  DGP model of gravity.  
While the sigma model violates positivity constraints imposed  by  
analyticity and  the Froissart bound,  the latter 
cannot be applied here because of  
the long-range tensor interactions that survive in the 
decoupling limit. We discuss further the properties of the 
Schwarzschild solution  that exhibits the 
gravitational mass-screening phenomenon.

\end{abstract}

\newpage

\section{Introduction and summary}

In a simplest perturbative expansion non-linear interactions  
in the DGP model \cite {DGP}  become strong  at a certain scale 
determined by the graviton mass, coupling constant and physical 
parameters of a  problem at hand (e.g., mass/energy of 
the source) \cite {DDGV}.  The strong  interactions and 
its scale  can be seen in a breakdown of the classical 
perturbation theory around a static source (in analogy 
with a similar phenomenon observed in Ref.  \cite {Arkady}
in the context of massive Fierz-Pauli gravity (FP) \cite {PF}).
More generically, this can be understood in terms of non-linear 
Feynman diagrams that are enhanced by inverse powers on the 
graviton mass \cite {DDGV}. This perturbative breakdown,
and  classical resummation of the corresponding large 
diagrams, is what makes the model agreable  with predictions 
of General Relativity \cite {DDGV}, \cite{Lue}-\cite{Kaloper}, 
and yields tiny but potentially measurable deviations 
from it \cite {DGZ,Lue1,Iorio}, \cite {GI,GI1}. 

It is instructive  to consider  the decoupling limit 
of the DGP model, as it was done in Refs. 
\cite {Ratt} and \cite {Ratt1}, motivated by the original  
studies of the decoupling limit of massive gravity in Ref. \cite {AGS}. 
It was argued that in this limit the DGP model 
reduces to a non-linear scalar model, 
while the self-interactions of a  tensor mode vanish \cite {Ratt,Ratt1}.
The remaining equation for the scalar field reads:
\beq
3\Box\pi= { (\Box\pi)^2 -  (\partial_\mu\partial_\nu\pi)^2\over 
\Lambda^3}\,,
\label{pieq0}
\eeq
where $\Lambda \equiv \mpl m_c^2 $ is held fixed while  
$\mpl\to \infty,m_c\to 0$, and $m_c$ plays the role of the graviton 
mass (the above equation is given here in a source-free region).  
Although, the results of \cite {Ratt,Ratt1} 
are similar in spirit to those found  in \cite {AGS} in the 
context of  massive gravity,
there are also important differences between the scalar 
models of Refs. \cite {AGS} and  \cite {Ratt,Ratt1}. 
These differences were shown to reproduce 
\cite {Creminelli,CRombouts}  the known  non-linear 
ghost-like instability  in 
the FP theory \cite {BD}, \cite {GGruzinov}, and were argued to be 
responsible for its absence  in (\ref {pieq0}) 
\cite {Creminelli,CRombouts}.

In spite of all the above, the formal similarity  between the 
two theories seems somewhat surprising
because of fundamental differences in their Lagrangians 
(see more on this in section 6). 
Most importantly, however, there exist an expression for the 
metric \cite {GI} such that: (a) It is an exact solution of the 
nonlinear DGP  equations on and near the brane; (b) It is a perfectly 
regular solution in the decoupling limit (see section 5);
(c) Yet, it does not satisfy Eq. (\ref {pieq0}).

The above observation  motivated us to re-examine 
the decoupling limit of the DGP model in the present work.
By analyzing the full set of nonlinear equations we will show below 
that one of the equations that survives in the decoupling limit 
is more general than (\ref {pieq0}), and looks as follows:
\beq
3\Box\pi+  {\t R}= { (\Box\pi)^2 -  (\partial_\mu\partial_\nu\pi)^2\over 
\Lambda^3}\,+ {{\t R}^2 - 3 {\t R}^2_{\mu\nu}\over 3 \Lambda^3} +
{{\t R}\Box \pi  - 2 {\t R}^{\mu\nu}\partial_\mu\partial_\nu\pi 
\over  \Lambda^3}\,. 
\label{pieq1}
\eeq
Here  ${\t R}\equiv \mpl R $ and $ {\t R}_{\mu\nu}\equiv 
\mpl {R}_{\mu\nu} $, where $R $ and ${R}_{\mu\nu}$ are the 
Ricci  scalar and tensor of a spin-2 field.
Eq. (\ref {pieq1})  reduces to Eq. (\ref {pieq0}) if, e.g., 
both Ricci tensor and scalar 
vanish.  However, this does not  have to be the case, as will be 
discussed below. For instance, the Schwarzschild solution of 
\cite {GI,GI1} gives  
${\t R}$ and components of ${\t R}_{\mu\nu}$
that are nonzero even in the decoupling limit, 
and satisfies (\ref {pieq1}), but not (\ref {pieq0}).

The issue of small fluctuations about classical backgrounds 
is different in (\ref {pieq0})  and (\ref {pieq1}).  
Eq. (\ref {pieq0})  implies that the fluctuations 
of the $\pi$ field are decoupled from those of the tensor 
and carry an independent physical meaning. On the other hand,  
Eq. (\ref {pieq1}) shows that the fluctuations of the 
$\pi$ field mix with those of a tensor. In fact, as we will see
below, the $\pi$ field can be completely absorbed into 
a tensor field, which survives as a nonlinearly interacting mode 
and propagates five physical polarizations.

More importantly, however, there exist  yet 
another non-linear equation that survives 
in the decoupling limit. The latter  is somewhat involved 
to be presented here without lengthy explanations (see 
Eq. ({\ref {sasaki2})).  Consistent solutions should satisfy that 
equation too. We will show that the spherically-symmetric 
solution of Eq. (\ref {pieq0}) found in Ref.
\cite {Ratt1}, although obeys (\ref {pieq1}),  
does not satisfy  Eq. (\ref {sasaki2}). Hence, it is not a solution of 
the DGP model\footnote{The solution of  
\cite {Ratt1}, in our nomenclature, is a solution of the 
equations of motion  that are obtained from  the sigma model Lagrangian 
in  which the $\pi$ field is decoupled from the tensor field (i.e., 
there are no mixings between them) \cite{Ratt,Ratt1}. 
In the solution discussed  in \cite{Gruzinov} the $\pi$ and 
tensor fields are sourcing each other.}. 
On the other hand, the solution of \cite {GI} satisfies both 
(\ref {pieq1}) and (\ref {sasaki2}).

As was recently shown in Ref. \cite {Ratt2}, the scalar model 
(\ref {pieq0}) presents a very interesting field-theoretic example:
In spite of its appearance as a local, Lorentz-invariant 
effective field theory, it exhibits certain hidden non-locality \cite {Ratt2}. 
The latter can manifest itself in superluminal propagation   
of perturbations on the background defined by the spherically-symmetric 
solution of \cite {Ratt1}. If the DGP model were to reduce 
in the decoupling limit to the scalar theory (\ref {pieq0}), then, 
any experimental test of this model would present a 
window of  opportunity for discovering small effects 
of an ${\cal O}(1)$ superluminality in gravitation \cite {Ratt2}. 

While this intriguing possibility can exist in the scalar 
sigma model \cite {Ratt2}, our results show that it is not offered by DGP.  
The solution  of  \cite {Ratt1}, and its fluctuations that manifest 
superluminal propagation, are not consistent 
solutions of the DGP model. More generally, 
the scalar model violates positivity constraints \cite {Ratt2} 
imposed  by analyticity and  the Froissart bound, however, 
the latter cannot be applied to the  present 
model of gravity  because of the long-range tensor interactions 
that survive in the decoupling limit.

The paper is organized as follows. In section 2 we discuss the ADM 
formalism for DGP and write down explicitly some of the key   
nonlinear equations. In section 3 we take the decoupling limit of 
those equations and derive Eq. (\ref {pieq1}).  In section 
4 we deal with  the remaining nonlinear equations 
of the model and find their decoupling limit. We also show that the 
solution of Ref. \cite {Ratt1} does not satisfy these equations.
In section 5 we discuss the physical interpretation of the 
Schwarzschild solution of Refs. \cite {GI,GI1} in the decoupling 
limit. We emphasize that this solution, although naively 
seems counterintuitive, actually  has a  clear physical 
interpretation in terms of the screening effects. In section 6, 
we discuss some interesting open questions.

\section{DGP in the  ADM formalism}
 
{}Let us start with the action of the DGP model \cite{DGP} 
in the ADM formalism \cite {ADM} (see, e.g., \cite {Dick}, 
\cite {CedricMourad}):
\beq 
{S}=\, {\mpl^2\over 2} \int d^4x\,dy \sqrt{ g} \left (  R~\delta(y)+
{m_c\over 2} 
\,N\,\left (R \,+\,K^2 -K_{\mu\nu}K^{\mu \nu} \right )\right )\,.
\label{ADMaction}
\eeq
Here, the $(4+1)$ coordinates are $x^M=(x^\mu,\ y)$, $\mu=0,
\dots, 3$;  $g$ and $R$ are the determinant and 4D 
curvature for the 4D components $g_{\mu\nu}(x,y)$ of the 5D metric
$g_{AB}(x,y)$.  $K=g^{\mu \nu}K_{\mu \nu}$ is the trace of the 
extrinsic curvature tensor defined as follows: 
\beq
K_{\mu \nu}\,=\,{1\over 2N}\,\left (\partial_y g_{\mu \nu} -\nabla_\mu 
N_\nu
-\nabla_\nu N_\mu \right )\,,
\label{K}
\eeq
and $\nabla_\mu $ is a covariant derivative w.r.t. the metric $g_{\mu \nu}$.
We introduced the {\it lapse} scalar field  $N$, and 
the {\it shift} vector field $N_\mu $ according to the standard 
rules:
\beq
g_{\mu y}\,\equiv \, N_\mu =g_{\mu\nu}N^\nu\,,~~~g_{yy}\,\equiv \,N^2\,+\,
g_{\mu\nu}\,N^\mu \,N^\nu\,. 
\label{adm}
\eeq 
Equations of motion of the theory are obtained  by varying 
the action (\ref {ADMaction}) w.r.t. $g_{\mu \nu}$, $N$ and  $N_\mu $. 
Here we start with a subset of two equations, the  
junction  condition across the brane, and the $\{yy\}$ 
equation that can be obtained by varying  the action 
w.r.t. $N$. The former reads as follows:
\beq\label{jc}
G_{\mu\nu} - m_c(K_{\mu\nu}-g_{\mu\nu}K)&=&T_{\mu\nu}/\mpl^2~,
\eeq
where $G_{\mu\nu}$ is the 4D Einstein tensor 
of the induced metric $g_{\mu\nu}$ and $T_{\mu\nu}$ is the 
matter stress tensor. The $\{yy\}$ equation  takes the form
\beq
R=  \,K^2 - K_{\mu \nu}^2 \,.
\label{yy}
\eeq
Note that (\ref {jc}) is valid only at $y=0^+$ while 
(\ref {yy}) should be fulfilled for arbitrary $y$.
The terms with the extrinsic curvature contain
derivatives w.r.t. the extra coordinate as well as
the $yy$  and $\mu y$  components of the metric.
Nevertheless, one can deduce  a single equation  that 
contains the 4D induced metric only!
This is done by expressing from (\ref {jc})
$K$ and $K_{\mu\nu}$ in terms of $R$ and $R_{\mu\nu}$ 
outside of the source 
\beq
K_{\mu\nu} ={R_{\mu\nu}-g_{\mu\nu}R/6\over m_c},~~~K={R\over 3m_c}\,,
\label{KKmunu}
\eeq
and substituting these expressions into (\ref {yy}). The result 
at $y=0^+$ is:
\beq
R&=&{R^2-3R^2_{\mu\nu}\over 3 m_c^2}\,.
\label{master}
\eeq
We will study the decoupling limit of this equation below.

\section{(De)coupling limit}

The purpose of this section is 
to show that the nonlinear interactions of the tensor  
field in the DGP model do not disappear in the decoupling limit.

The decoupling limit in DGP is defined as follows \cite {Ratt}:
\beq
\mpl\to \infty,~~m_c\to 0,~~~M\to \infty\,,
\label{lim}
\eeq
where $M$ is a mass of a source entering the stress-tensor 
$T$ and the following quantities are held finite and fixed
\beq
\Lambda \equiv (\mpl m_c^2)^{1/3},~~~{M\over \mpl}\,.
\label{fixed}
\eeq
To take this limit in (\ref {master}), we multiply both sides
of that equation  by $\mpl$ and obtain
\beq
{\bar R}&=&{{\bar R}^2-3{\bar R}^2_{\mu\nu}\over 3 \Lambda^3}\,,
\label{masterBar}
\eeq
where we defined ${\bar R}\equiv \mpl R $ and $ {\bar R}_{\mu\nu}\equiv 
\mpl {R}_{\mu\nu} $.  Now we are ready to  proceed to 
(\ref {lim},\ref {fixed}). Before doing so, 
it is instructive to make 
parallels with FP gravity. The analog of the mass term in the 
DGP action (\ref {ADMaction}) is the one containing  squares
of the extrinsic curvature.  Looking at the expression for the 
extrinsic curvature (\ref {K}) 
one can think of  $g_{\mu\nu}$ in DGP to be an analog of   
the tensor field of FP gravity in which the St\"uckelberg
field is manifestly exposed in the mass term \cite {AGS}. Then, 
$N_\mu$ is an analog of the vector-like St\"uckelberg field
as far as the  reparametrizations with the gauge-function 
$\zeta_A(x,y)=(\zeta_\mu(x,y),0)$ are concerned. The longitudinal 
component of $N_\mu$ should play  the role similar to that of the 
longitudinal component of the vector-like St\"uckelberg field
of FP gravity (note that there is an additional St\"uckelberg
field $N$ in DGP, the important role of which will
be discussed in section 6).

Equation (\ref {masterBar}), on the other hand,
does not contain any of the St\"uckelbeg fields
but  the tensor $g_{\mu\nu}$ which in the decoupling limit has a 
conventional scaling\footnote{If we were to do perturbative calculations, 
this would imply the gauge choice for which the 
St\"uckelberg fields are manifestly present.}. 
That is why it is straightforward to take the 
limit directly in (\ref {masterBar}). For this we recall that 
\beq
{\bar R}&=& \Box {\bar h} - \partial^\mu \partial^\nu {\bar h}_{\mu\nu}
+{\cal O}\left ( {\bar h}\Box {\bar h} \over \mpl \right )\,,\nonumber \\
{\bar R}_{\mu\nu}&=&{1\over 2} \left ( \Box {\bar h}_{\mu\nu} 
- \partial_\mu \partial^\alpha {\bar h}_{\alpha \nu}
- \partial_\nu \partial^\alpha {\bar h}_{\alpha \mu} +
\partial_\mu \partial_\nu {\bar h} \right )+ 
{\cal O}\left ( {{\bar h}\Box {\bar h} \over \mpl} \right )\,,
\label{Rs}
\eeq
where we defined a field 
${\bar h}_{\mu\nu}\equiv \mpl (g_{\mu\nu}-\eta_{\mu\nu})$,
which has the canonical dimensionality and 
is held fixed in the limit\footnote{The signs in the expressions
(\ref {Rs}) and the definition of ${\bar h}_{\mu\nu}$ given above 
determine our convention for the sign of the curvature tensor.}.

Note that we have  not done any small field approximation,
but merely expanded the nonlinear expressions for  ${\bar R}$ and 
${\bar R}_{\mu\nu}$ in powers of ${\bar h}$  and 
took  the decoupling limit.  Because all the  nonlinear terms 
in ${\bar R}$ and  ${\bar R}_{\mu\nu}$  are suppressed by extra 
powers of $\mpl$, the curvatures reduce to their linearized form.

The expressions in (\ref {Rs}), should be substituted into (\ref {masterBar}). 
To make closer contact with the $\pi$ language 
of Section 1 we perform the following shift of the 
${\bar h}_{\mu\nu}$ field
\beq
{\bar h}_{\mu\nu}= {\t h}_{\mu\nu} +\eta_{\mu\nu} \pi\,,
\label{shiftBar}
\eeq
where the $\pi$ field has a canonical dimensionality and 
is held fixed in the decoupling limit. With this substitution
Eq. (\ref {masterBar}) turns into (\ref {pieq1}) which we 
repeat here for convenience
\beq
3\Box\pi+ {\t R} = { (\Box\pi)^2 -  (\partial_\mu\partial_\nu\pi)^2\over 
\Lambda^3}\,+ {{\t R}^2 - 3 {\t R}^2_{\mu\nu}\over 3 \Lambda^3} +
{{\t R}\Box \pi  - 2 {\t R}^{\mu\nu}\partial_\mu\partial_\nu\pi 
\over  \Lambda^3}\,. 
\label{pieqExp}
\eeq
$\t R$ and ${\t R}_{\mu\nu}$ denote the linear
terms on the right-hand-sides of (\ref {Rs}) where
${\bar h}_{\mu\nu}$ is replaced by ${\t h}_{\mu\nu}$.

Eq. (\ref {pieqExp}) shows 
that a tensor field would have non-linear interactions 
in the decoupling limit as long as there are no additional 
equations constraining  all the components 
of ${\t R}_{\mu\nu}$ to be zero.  We will turn to the remaining 
equations in the next section and show that they do not necessarily imply 
vanishing ${\t R}_{\mu\nu}$. Before that we would like to make a comment
on small fluctuations about classical solutions. The $\pi$ field 
in (\ref {pieqExp}) can be reabsorbed back into the tensor field 
${\bar h}_{\mu\nu}$ using (\ref {shiftBar}). Small  
perturbations on any  background in the decoupling limit are 
those of ${\bar h}_{\mu\nu}$. The given background  is  
what defines the light-cone, and there are no additional degrees of 
freedom that could propagate outside of that light-cone.

\section{More bulk equations}

Some of  the bulk equations have not been considered 
in our discussions so far. These are the $\{\mu y\}$ and 
bulk $\{\mu\nu\}$ equations (the $\{\mu\nu\}$ equation on 
the brane (\ref {jc}) has already been taken into account).
We will discuss them in the present section.

We start with the  $\{\mu y\}$ equation which for arbitrary $y$ 
reads as follows:
\beq
\nabla_\mu K= \nabla^\nu K_{\mu\nu}\,.
\label{mu5}
\eeq
The covariant derivative in the above equation 
is the one for $g_{\mu\nu}$. At $y=0$ Eq. (\ref {mu5})  is 
trivially satisfied due to  (\ref {KKmunu}).  For $y\neq 0$, 
(\ref {mu5}) sets the relation between $N_\mu$, $N$ and $g_{\mu\nu}$.  
Hence, (\ref {mu5}) gives a relation between these 
quantities for both $y=0$ and $y\neq 0$.

One can use the bulk $\{yy\}$ equation (\ref {yy}) to determine 
$N$ in terms of $N_\mu$ and $g_{\mu\nu}$, and then use 
(\ref {mu5}) in order to express $N_{\mu}$ in terms of 
$g_{\mu\nu}$.  If so, there must exist one more equation which 
should allow to determine the bulk $g_{\mu\nu}$ itself. This is 
the bulk $\{\mu\nu\}$ equation, to which we turn now.
The latter can be written in a few different 
ways. For the case at hand, the formalism by Shiromizu, Maeda and Sasaki 
\cite {Sasaki} is most convenient.
In this approach, the bulk $\{\mu\nu\}$ equation and the junction 
condition can be combined to yield a  $\{\mu\nu\}$ equation 
at $y\to 0^+$. This gives a ``projection''
of the bulk $\{\mu\nu\}$ equation onto the brane. Since the bulk 
itself is empty in our case, 
this is the most restrictive form one can work with. 
The equation reads as follows \cite {Sasaki}:
\beq
G_{\mu\nu} = K K_{\mu\nu} - K_\mu^\rho K_{\nu\rho} -{1\over 2}
g_{\mu\nu} (K^2 -K_{\mu\nu}^2) - E_{\mu\nu}\,,
\label{sasaki1}
\eeq
where all the quantities are taken at $y=0^+$.
$G_{\mu\nu}$ denotes the 4D Einstein tensor of the metric $g_{\mu\nu}$, 
$E_{\mu\nu}$ denotes the electric part of the bulk Weyl tensor, 
projected onto the brane. An important property of the Weyl 
tensor is that it is invariant under the conformal 
transformations. Moreover, $E_{\mu\nu}$ is 
traceless.

We now turn to the decoupling limit in this equation. 
For this we multiply both sides of (\ref {sasaki1}) by $\mpl$,
exclude $K$ and $K_{\mu\nu}$ using (\ref {KKmunu}), and take the 
limit (\ref {lim}, \ref {fixed}) holding the canonically normalized  
components of $g_{\mu\nu}$ fixed. The resulting equation reads:
\beq
{\bar G}_{\mu\nu} = {{\bar B}_{\mu\nu}\over \Lambda^3} - {\bar E}_{\mu\nu}\,,
\label{sasaki2}
\eeq
where 
\beq
{\bar B}_{\mu\nu}\equiv -  {\bar G}_{\mu\alpha}{\bar G}^{\alpha}_\nu +
{1\over 3}{\bar G}^{\alpha}_\alpha   {\bar G}_{\mu\nu}
+{1\over 2} \eta_{\mu\nu} {\bar G}_{\alpha \beta}{\bar G}^{\alpha\beta} 
-{1\over 6} \eta_{\mu\nu} ({\bar G}^{\alpha}_\alpha)^2\,,  
\label{B}
\eeq
and ${\bar G}_{\mu\nu}\equiv \mpl G_{\mu\nu}$,  
${\bar E}_{\mu\nu}\equiv \mpl E_{\mu\nu}$.
This is the equation that has to be satisfied
by any consistent solution.  

It is straightforward  to check that  the trace of the above  equation is 
nothing  but (\ref {masterBar}). Moreover, as in Eq. (\ref {masterBar}), 
there are nonlinear interactions of the tensor field 
that survive in the decoupling limit in (\ref {sasaki2}).  This is 
true  irrespective of the explicit form of  ${\bar E}$, 
as long as it's traceless. Once the values of $N$ and $N_\nu$ are 
determined, as described at the beginning of this section, one can 
calculate $E_{\mu\nu}$ in terms of $g_{\mu\nu}$. Thus, 
Eq. (\ref {sasaki2}) turns into a single equation for the 
determination of $g_{\mu\nu}$.  

The above considerations show that the DGP model in general 
does not reduce to the scalar theory (\ref {pieq0}) in the 
limit (\ref {lim},\ref {fixed}). 
This can also be deduced  by looking  at the  solutions 
of the scalar model (\ref {pieq0}) which are not solutions of 
(\ref {sasaki2}).  One example of this is the spherically-symmetric 
static solution of (\ref {pieq0}) found in \cite {Ratt1}.  We will discuss 
this solution in the remaining part of this section. 
The ansatz  corresponding to the solution of 
\cite {Ratt1} reads:
\beq
{\tilde R}_{\mu\nu} ({\t h}) =0,~~~ 
{\bar G}_{\mu\nu}({\bar h})= \partial_{\mu} \partial_{\nu} \pi- \eta_{\mu\nu}
\Box \pi,~~~
K_{\mu\nu} = {m_c\over \Lambda^3}\partial_{\mu} \partial_{\nu} \pi\,.
\label{ansatz}
\eeq
One can calculate  the expression for ${\bar E}_{\mu\nu}$ on 
the ansatz (\ref {ansatz}) (${\bar E}_{\mu\nu}$ comes out to be non-zero), 
and use that expression in  (\ref {sasaki2}). The resulting  equation, 
after substitution  (\ref {pieq0}), reduces to Eq. (\ref {newpi}) 
presented below. These calculations are tedious, and instead of them,  
we present here relatively easier calculations 
that produce the same result (\ref {newpi}) (this also serves as a 
self-consistency check of our result). 

Since the  solution (\ref {ansatz}) is written in terms of the 
extrinsic curvature, it is easier  to check its compatibility 
with the bulk $\{\mu\nu\}$ equation also written in terms of 
this quantity. This equation takes the form \cite {CedricMourad}\footnote{
Note that our curvature sign conventions are different from those of Ref. 
\cite {CedricMourad}.}:
\beq
&G_{\mu\nu}&={1\over 2} g_{\mu\nu} \left(K^2 -K_{\alpha\beta}^2\right )
+2 \left ( K_\mu^\alpha K_{\nu\alpha} -KK_{\mu\nu} \right ) \nonumber \\
&-&{2\over N}   \left ( \nabla_\nu(N_\mu K) -\nabla_\alpha (K_\nu^\alpha 
N_\mu) -{1\over 2} g_{\mu\nu} \nabla^\alpha (KN_\alpha) +{1\over 2}
\nabla^\alpha(N_\alpha K_{\mu\nu}) \right ) \nonumber \\
&-& {\nabla_\mu \nabla_\nu N -g_{\mu\nu}\nabla^2 N\over N}
+g_{\mu\alpha} g_{\nu\beta} {\partial_y (\sqrt{g} (
K^{\alpha\beta} - Kg^{\alpha\beta} ))  \over N\sqrt{g}  }\,.
\label{bulkKmunu}
\eeq
To check whether the solution of \cite {Ratt1} is consistent
with (\ref {bulkKmunu}), we take the decoupling limit in the 
above equation with the substitution \cite {Ratt,Ratt1}
\beq
h_{\mu y} = - {m_c\over \Lambda^3} \partial_\mu \Pi,~~~
h_{yy} = - {2m_c\over \Lambda^3} \partial_y \Pi,~~~\Pi(x,y) 
= e^{-y \sqrt{-\Box}} \pi(x)\,.
\label{Rattans}
\eeq
This can be expressed in terms of the lapse scalar and shift vector 
in the decoupling limit
\beq
N_\mu=h_{\mu y}, ~~~N^2 = g_{yy}+ {m_c^2\over \Lambda^6}(\partial_\mu 
\Pi)^2\,.
\label{nnh}
\eeq
The decoupling limit for $g_{yy}$ has been specified 
in Ref. \cite {Ratt1} only in the leading approximation
$g_{yy}= 1- 2 m_c \partial_y \Pi/\Lambda^3$. However, for  consistency,
the subleading term in (\ref {bulkKmunu}}) is also needed, 
otherwise the ansatz itself would be 
inconsistent as it would  violate some 
sacred properties of the bulk equation (\ref {bulkKmunu})
(for instance, a non-linearly incomplete ansatz, after substitution into 
the bulk equation, does  not respect the Bianchi identities and 
would not reproduce correctly the  trace equation). On the other hand, 
in the next-to-leading order there is a unique 
non-linear completion for $g_{yy}$ that  is consistent
with the Bianchi identities and trace equation. 
The latter reads $g_{yy} = 1 - 2 m_c \partial_y \Pi/\Lambda^3
+m_c^2 (\partial_y\Pi)^2/\Lambda^6$. For this ansatz, the 
expression for $N$ in the decoupling limit reads as follows: 
$N\simeq 1- m_c \partial_y \Pi/\Lambda^3
+m_c^2 (\partial_\mu\Pi)^2/2\Lambda^6$. The latter 
combined with  the other components above gives a consistent ansatz on 
which Eq. (\ref {bulkKmunu}) looks as follows:
\beq
\partial_{\mu} \partial_{\nu} \pi- \eta_{\mu\nu}\Box \pi =
-\eta_{\mu\nu} { (\Box\pi)^2 -  (\partial_\alpha\partial_\beta\pi)^2\over 
2 \Lambda^3} + { (\Box\pi)\partial_\mu\partial_\nu\pi   -  
\partial_\mu\partial_\alpha \pi \partial_\nu\partial^\alpha \pi \over 
\Lambda^3}\,.
\label{newpi}
\eeq
As before, trace of (\ref {newpi}) gives precisely  (\ref 
{pieq0}), i.e., the former equation  is more general than (\ref {pieq0}).
Hence, not all the solutions of (\ref {pieq0}) satisfy (\ref {newpi}).
For instance, consider the $\{0 0\}$ component of 
Eq. (\ref {newpi}) and look at a static spherically-symmetric solution.
Eq. (\ref {newpi}) reduces to
\beq
\Delta  \pi ={ (\Delta \pi)^2 -  (\partial_i \partial_j\pi)^2\over 
2 \Lambda^3}\,,
\label{00}
\eeq
where $\Delta$ denotes a 3D Laplacian and $i,j=1,2,3$.
The above equations is incompatible with the time independent 
part of (\ref {pieq0}), unless both left and 
right hand sides  of (\ref {pieq0}) and (\ref {00}) are zero.
The solution found in \cite {Ratt1} does not satisfy this 
constraint, and, therefore, it does not satisfy the $\{0 0\}$ 
component of Eq. (\ref {newpi}).  

Likewise, one can show that  the solution of \cite {Ratt1} 
does not satisfy the $\{i j\}$ components of (\ref {newpi}) either. 
To demonstrate this, we consider the $\mu\neq \nu$ part of 
Eq. (\ref {newpi}) and look at a spherically-symmetric static 
solution.  Using  the notations of \cite {Ratt1}, 
$E_j = \partial_j\pi \equiv r_jE/r$, where $j=1,2,3$, 
equation (\ref {newpi}) with $i \neq j$ gives
\beq
\left ( {d E\over dr} -{E\over r}\right ) 
\left (1 - {E\over r \Lambda^3}\right ) = 0\,. 
\label{eqE}
\eeq
The only  nontrivial solution of (\ref {eqE}) reads: 
$E=c r$, where $c$ can  be an arbitrary 
constant (including $c=\Lambda^3$), which is not the  solution 
of \cite {Ratt1}.  This is because the ansatz (\ref {Rattans})
is a pure gauge in the bulk only in the linearized approximation,
but not in a full non-linear theory. In the next section we discuss a 
solution that exactly satisfies the above equations on and near 
the brane \cite {GI,GI1}.

\section{Schwarzschild solution}

The Schwarzschild solution found in \cite {GI} satisfies all 
the above nonlinear equations on and near the brane. 
Here, we will discuss some of the main features of this solution in the 
decoupling limit and give its further physical interpretation.
  
For a static point-like source, $T_{00}=-M\delta^3(\vec{x})$
with $T_{ij}=0$. A new physical scale emerges in this problem
as a combination of $r_c$ and $r_M$ \cite {DDGV} 
($r_M\equiv 2G_NM$  is the  Schwarzschild  radius 
and $G_N$ the Newton constant)
\beq
r_*\equiv (r_M r_c^2)^{1/3}\,.
\label{r*}
\eeq
(This is similar to the Vainshtein scale in massive gravity
\cite {Arkady}). It is straightforward to check that
this scale is finite and fixed in the decoupling limit 
(\ref {lim},\ref {fixed}).

The  metric on/near the brane was found  
exactly in Refs. \cite {GI,GI1}. Here, for our purposes 
it suffices to concentrate  on  the  Newton  potential 
alone $\phi(r)=h_{00}/2$.  In the notations adopted in the 
previous sections  
\beq
g_{00}=-1+h_{00}= -1+{\bar h}_{00}/\mpl\,.
\label{phi}
\eeq
The exact expression for ${\bar h}_{00}$ \cite {GI}
is a solution of (\ref{master}).
Furthermore, (\ref {sasaki1}) can be used to determine
the off-diagonal and $\{yy\}$ components of the metric near the 
brane, as was done in \cite {GI,GI1}. Below we will check 
directly that the terms containing curvatures in (\ref {pieq1})
are not zero on this solution. Hence, it is physically different
from the solution of \cite {Ratt1}. 

For simplicity  we concentrate on  the  $r<< r_*$ region and 
do the following: we split  $\bar h$  into two parts  according to 
(\ref {shiftBar}),  and for comparison with 
\cite {Ratt1} insist that $\pi$ has the form obtained in 
\cite {Ratt1}. Using the results of \cite {GI}, for scales  
$r\ll r_*$, we get:
\beq
{{\t h}_{00}\over \mpl} = {r_M\over r}- \sqrt {2} 
m_c^2r^2 \left ({r_*\over r}\right )^{3\over 2} \left [
{\alpha \over \sqrt {2}} \left({r\over r_*}\right)^{\beta } - 1 \right ]
\label{potential}
\eeq
where $\beta = 3/2 - 2(\sqrt{3}-1)\simeq 0.04$, and      
$\alpha \simeq \pm 0.84$. While the $\pi$ field takes the form:
\beq
{\pi \over \mpl}=\sqrt{2} m_c^2r^2 \left({r_*\over r}\right)^{{3\over 2}}~.
\label{pisol}
\eeq
Note that  both ${\t h}_{00}$ and $\pi$ given above 
are finite in the limit (\ref {lim},\ref {fixed}).
It is the non-trivial part (\ref {potential}) that differentiates the 
solution of \cite {GI,GI1} from that of \cite {Ratt1}.
In particular, the curvatures $\t R$ in Eq. (\ref {pieq1})
are nonzero on this solution.

The solution of \cite {GI,GI1} has a number of interesting physical 
properties. Here we'd like to emphasize  the effect of the 
gravitational mass screening \cite {GI,GI1}. 
In the region $r\ll r_*$ the solution  recovers the results 
of the  Einstein theory with tiny, but potentially 
measurable deviations \cite {DGZ,Lue1,GI1}.  
Gravitational screening, on the other hand, manifests itself 
for $r>>r_*$.  In this region, the solution (Newton's potential) 
scales as $\sim r_*r_M/r^2$. Naively this seems  counterintuitive, 
since it contradicts  the $1/r$ scaling expected from perturbation 
theory in the region $r_*\ll r \ll r_c$ \cite {DGP}. 
However, as was explained in \cite {GI,GI1}, 
there is no actual contradiction since the perturbative approach does 
not take into account the  effect of gravitational mass screening 
of the source.  Unlike in conventional GR, a static source
in this theory produces a non-zero scalar curvature  even outside 
of the source \cite {GI}.  This curvature extends to scales $\sim r_*$.
Hence, the source is surrounded by a huge halo of scalar curvature. 
This halo screens the bare mass of the source. It is easy to estimate 
that the screening  mass should be of the order of the bare 
mass itself \cite {GI1}: A deviation from the conventional GR metric 
at  $r\ll r_*$ scales as $m_c \sqrt{r_M r}$ (we ignore small $\beta$ here.)
This can give rise to the scalar curvature that scales as  
$m_c\sqrt{r_M}r^{-3/2}$. The curvature extends approximately 
to  distances $r \sim r_*$. Because of this 
the integrated curvature scales as  $m_c \sqrt{r_M} r_*^{3/2}\sim r_M$, and 
the "effective mass" due to this curvature can be estimated to be 
$r_M \mpl^2 \sim M$, i.e., of the order of the bare mass $M$ itself!

Given the above  estimate, there could be either  complete or 
almost complete  ``gravitational mass screening'' for the problem at hand.  
Which one is realized in DGP? As we will argue below, 
the solution found in  \cite {GI,GI1} suggests that the screening 
is complete from the 4D point of view and is incomplete
from the 5D perspective. Let us discuss this in more detail. 
From a 4D perspective (i.e., from the point of view of the induced 
metric on the brane) 
the solution  behaves as a spherically-symmetric distributions of
mass/energy of  radius $r\sim r_*$, with the bare  mass $M$ 
placed in the center, and the screening  halo surrounding it. 
The solution \cite {GI} shows that the potential at $r\ge r_*$ has 
no ``monopole moment'', i.e., it has no $1/r$ scaling. 
Instead, it has the  ``dipole moment'' that scales as $\sim r_* r_M/r^2$, 
which fits the interpretation of a potential due to a 4D 
spherically-symmetric mass distribution of size 
$\sim r_*$ and zero net mass\footnote{We thank Gia Dvali for pointing out  
the dipole analogy to us. Notice, also that there is no 4D Birkhoff's 
theorem in this case.}. Hence, the 4D screening should be  
complete. The picture  is slightly different from the  5D 
perspective. In the definition
of the 5D ADM mass the off-diagonal component  of the 5D 
metric of \cite {GI,GI1}  is also entering (this  component vanishes 
on the brane  and that is why it does not contribute to the 4D ADM 
mass).  Because of this component, there is no complete screening of 
the 5D ADM mass, and  the potential has the 5D ``monopole'' component
which scales as $\sim r_*r_M/r^2$ and yields the 5D ADM mass
$\sim M(r_M/r_c)^{1/3}$ \cite {GI}. The potential due to an  
incompletely  screened 5D ``monopole'' smoothly 
matches onto the 4D potential due to a 4D ``dipole''.

\section{Discussions and outlook}

It looks like the present model works in subtle ways.
Yet, there are a number of issues that still need to be 
understood. We will outline some of them below.

{\it (1). DGP vs. massive gravity}.  
The results of the present work show that in 
the decoupling limit massive gravity and 
the DGP model behave very differently. It would be instructive  
to understand  this  in terms of the Lagrangians
of the two theories. It is useful to start by recalling  some 
details of the decoupling limit in 4D massive gravity \cite {AGS}.
The most convenient way is to use the St\"ukelberg method and 
complete the FP mass term to a reparametrization  invariant
form \cite {AGS}. This can be done order-by-order in powers of 
the fields.  Let us call the 4D massive gravitational field 
$f_{\mu\nu}$, the corresponding St\"uckelberg vector field $A_\mu$ 
and its longitudinal part $\phi$.  Due to the  mass term the 
$\phi$ field acquires  a  kinetic mixing term with the $f$ 
field. The quadratic part of the action can be diagonalized by a conformal 
transformation  $f_{\mu\nu} ={\t f}_{\mu\nu}+\eta_{\mu\nu}\phi$.  
As a result, the 
$\phi$ field acquires its own kinetic term as well as coupling  to 
the trace of the stress-tensor. 
The only nonlinear interactions that survive in the decoupling limit 
are those of $\phi$, all the nonlinear terms containing 
$\t f$ vanish \cite {AGS}.  The analog of the $f_{\mu\nu}$ field in DGP is 
$g_{\mu\nu}$, and the analog of $A_\mu$ is $N_\mu$, as it can be 
read off (\ref {ADMaction}). Besides these fields  there is 
an additional field $N$. In the 
linearized theory it enters the action (\ref {ADMaction}) 
as a Lagrange multiplier and enforces a  constraint  that is 
consistent with the linearized Bianchi  identities. A 
linear combination of this field with the longitudinal 
component of $N_\mu$ is an analog  of  the longitudinal component of 
the $A_\mu$  field of massive gravity -- these  modes acquire their  
own kinetic terms through mixing with the tensor fields. 
However, the similarities  end here. 
At the nonlinear level the $N$ field ceases to 
be a Lagrange multiplier, but enters  (\ref {ADMaction}) 
algebraically. Hence, it can be integrated out explicitly. 
The resulting action is a functional of
$g_{\mu\nu}$ and $N_\mu$ only, however, it is very different from 
the action of massive gravity expressed in terms of $f_{\mu\nu}$ and $A_\mu$.
The former contains non-local interactions between 
$g_{\mu\nu}$ and $N_\mu$. Moreover, if expanded on a flat background, 
it stays non-local  and does not produce quadratic mixing terms
between them. Under the circumstances, a way to proceed is to solve 
first equations for a  nontrivial background, then  expand the 
non-local action in perturbations about the background, and only then take  
the limit. This is, essentially,  what we have done  in the present 
work, but in easier terms of the equations  of motion.

{\it (2). Analyticity and the unitarity bound}.
The tree-level one-particle exchange 
amplitude in DGP is nonlocal from 4D point of view as it contains terms 
$(\Box + m_c \sqrt{-\Box})^{-1}$ \cite {DGS}. 
The amplitude has a branch-cut because of the square root.
As a result, there is a way do define a contour in the 
complex plane  so that the pole in the amplitude 
ends up being on the second Riemann  sheet (see, e.g.,\cite {GGIan}). 
The usual 4D dispersion  relation can be written for this 
amplitude, and it does  respect  4D analyticity. Having this 
established, one can consistently take $m_c\to 0$, as it is done in the 
decoupling limit.  How about the non-linear amplitudes? 
Generically, those contain terms that are  singular
in the $m_c\to 0$ limit (e.g., different powers of $1/m_c\sqrt{-\Box}$)
\cite {DDGV}.  These are the terms that make 
the perturbative expansion to break down at the scale $r_*$.
The results of the present work suggest that it is unlikely 
that the limiting theory below $r_*$ can be thought of a local 4D 
theory of massless  helicity-2, helicity-1 and helicity-0 states.  
However, in the  regions and/or in the  backgrounds 
where the conventional perturbative expansion 
is valid (i.e, where the above singular terms can be ignored), 
it is reasonable to expect that 4D analyticity is respected.

What is more certain, however, is  that the Froissart bound cannot be 
assumed to hold even for the limiting theory.
The reason being that in the decoupling limit
there are long-range tensor interactions present, and 
the Froissart saturation can only be taking place if 
such interactions were absent. Clearly, the present theory of 
gravity has no mass-gap.

{\it (3). (Important) miscellanea}.
The issue of the UV completion and quantum consistency 
of the model need more detailed studies (see, e.g., \cite {Dvali}). 
In this respect, it would be interesting 
to pursue furtherer the issue  of  the string theory realization of 
brane induced gravity, perhaps, along the lines of Refs. 
\cite {Kiritsis,Ignatios,Eman}.

Finally, in the present work we have not discussed the selfaccelerated 
universe \cite {Deffayet,DDG}. On issues of 
consistency of this approach will be reported in \cite {DGI}.

%

\section*{Acknowledgments}

{} We would like to thank C\'edric Deffayet, Gia Dvali and Nemanja Kaloper   
for useful discussions and communications.  The work of both of us was 
supported in part by  NASA Grant NNGG05GH34G, and in part 
by NSF Grant PHY-0403005.


\end{document}